\documentclass[prb,twocolumn,superscriptaddress,showpacs]{revtex4}
\usepackage{times}
\usepackage{amsmath}
\usepackage{amsfonts}
\usepackage{graphicx}
\usepackage{pst-all}

\begin{document}

\title{Fractional Quantum Hall Hierarchy and the Second Landau Level}
\author{Parsa Bonderson}
\affiliation{Microsoft Research, Station Q, Elings Hall, University of California,
Santa Barbara, CA 93106}
\affiliation{California Institute of Technology, Pasadena, CA 91125}
\author{J. K. Slingerland}
\affiliation{Dublin Institute for Advanced Studies, School for Theoretical Physics, 10 Burlington Rd, Dublin 4, Ireland}
\affiliation{Department of Mathematical Physics, National University of Ireland, Maynooth, Ireland}
\affiliation{Department of Physics, University of California, Riverside, CA 92521}
\affiliation{California Institute of Technology, Pasadena, CA 91125}
\date{\today}

\begin{abstract}
We generalize the fractional quantum Hall hierarchy picture to apply to arbitrary, possibly non-Abelian, fractional quantum Hall states. Applying this to the $\nu = 5/2$ Moore--Read state, we construct new explicit trial wavefunctions to describe the fractional quantum Hall effect in the second Landau level. The resulting hierarchy of states, which reproduces the filling fractions of all observed Hall conductance plateaus in the second Landau level, is characterized by electron pairing in the ground state and an excitation spectrum that includes non-Abelian anyons of the Ising type. We propose this as a unifying picture in which $p$-wave pairing characterizes the fractional quantum Hall effect in the second Landau level.
\end{abstract}

\maketitle










\section{Introduction}

The fractional quantum Hall (FQH) plateaus in the first Landau level are described rather well by the Laughlin states~\cite{Laughlin83} and the Abelian hierarchy states constructed over them~\cite{Haldane83,Halperin84}. The observed filling fractions, the measured fractional charge of quasiparticles~\cite{Goldman95,Saminadayar97,Picciotto97}, and recent results from interferometric experiments~\cite{Goldman05b,Goldman07} all support this picture, and are backed up by a wealth of theoretical and numerical evidence. The physics of the second Landau level, however, remains far more perplexing, with the prominence of an even-denominator $\nu=5/2$ FQH state~\cite{Willett87,Pan99} that cannot be explained by the standard hierarchy. Fully developed FQH plateaus have been observed at $\nu=7/3$, $12/5$, $5/2$, $8/3$, and $14/5$~\cite{Pan99,Xia04}, but advances in experiments and sample quality may find additional plateaus developing where ``features'' have been observed, including at another intriguing even-denominator $\nu=19/8$~\cite{Xia04,Pan08}. There are also sometimes observed signs of what appear to be developing FQH states at $\nu = 11/5$, $16/7$, $13/5$, $19/7$, and $25/9$, but these dematerialize as lower temperatures are attained~\cite{Xia04,Pan08}.

The currently held view is that the $\nu=5/2$ plateau is characterized by $p$-wave pairing of composite fermions and described by the Moore--Read (MR) state, which gives the dramatic prediction of quasiparticles with non-Abelian braiding statistics~\cite{Moore91,Nayak96c}. Numerical studies support the case for the MR state (and its particle-hole conjugate)~\cite{Morf98,Rezayi00,Feiguin08a,Wan08a,Peterson08,Feiguin08b}. Recent tunneling shot-noise experiments have confirmed the $e/4$ fundamental quasihole charge of the $\nu=5/2$ state expected for the MR state~\cite{Dolev08}. Further recent evidence from the scaling behavior in dc transport experiments~\cite{Radu08} at $\nu=5/2$ best agrees with the particle-hole conjugate of MR~\cite{Lee07,Levin07}.

Although the remaining observed filling fractions in the second Landau level have odd denominators, numerics indicate that the electron correlations for $7/3 \leq \nu \leq 8/3$ have a non-Laughlin character similar to that of $\nu=5/2$~\cite{Wojs01}, and so only $\nu=14/5$ is expected to be an Abelian state. Aside from the Abelian hierarchy states, the non-Abelian Read--Rezayi (RR) $k$-body clustered states~\cite{Read99} (which include MR) and their particle-hole conjugates are essentially the only single layer, spin-polarized descriptions that have been proposed for these FQH plateaus. There are also other proposed non-Abelian states, such as the $\text{SU}\left(N\right)_{k}$ NAF (non-Abelian FQH) states~\cite{Wen91a,Blok92}, whose potential relevance to the second Landau level is largely unexplored. The experiments of Ref.~\onlinecite{Dolev08} also found $e/3$ fundamental quasihole charge for the $\nu=7/3$ and $8/3$ states, which rules out their being RR or NAF states (as these candidate states have $e/6$ fundamental quasihole charge). To our knowledge, there have been no non-Abelian states with $e/3$ fundamental quasihole charge proposed for the $\nu=7/3$ and $8/3$ plateaus (prior to this paper).

Clearly it is important to fully explore the possibilities of non-Abelian candidate FQH states that are consistent with the experimental status of the second Landau level, or, in fact, to have \emph{any} such candidates whatsoever for the observed second Landau level filling fractions, especially for those in the range $7/3 \leq \nu \leq 8/3$ where non-Abelian states are expected to occur. Motivated by this, we generalize the hierarchy construction so that it can also be applied to non-Abelian FQH states. By applying this generalized hierarchical scheme to the MR state, we produce new $p$-wave paired states and corresponding explicit wavefunctions. Remarkably, these states occur at \emph{all} the observed second Landau level filling fractions (and even weaker observed features), and in natural order for those expected to be non-Abelian. This leads to a simple, intuitive picture in which $p$-wave pairing is the characteristic property of the second Landau level.

In our application of hierarchy to the MR state, the fundamental quasiholes/quasielectrons are assumed to form a gas of bound pairs, which is projected onto wavefunctions analogous to the first Landau level hierarchy. This generates FQH states with the conformal field theory (CFT) structure $\left. \text{Ising}\times \text{U}\left( 1\right) _{K}\right| _{\mathcal{C}}$, where $K$ is a coupling constant matrix for a hierarchy in the charge sector of the MR state (i.e. starting from  $K_{00}=2$), and $\mathcal{C}$ is the topological charge spectrum. The product structure of the underlying CFT is clearly reflected in the resulting trial wavefunctions; see Eqs.~(\ref{eq:extwothirds},\ref{eq:extwofifths},\ref{eq:BSCF1},\ref{eq:BSCF3}) for some of the more relevant examples. In fact, some of the trial wavefunctions are simply a product of a Pfaffian and a bosonic hierarchy wavefunction [e.g. Eq.~(\ref{eq:extwofifths})], or a product of a bosonic MR wavefunction and a composite fermion wavefunction [e.g. Eq.~(\ref{eq:BSCFapprox})].

The paper is structured as follows: In Section II, we provide a generalization of the hierarchy scheme that can be applied to non-Abelian FQH states. In Section III, we provide a description of particle-hole conjugation and the Haldane--Halperin (HH) hierarchy that will be helpful in generalizations. In Section IV, we apply the hierarchy methods to the MR state, producing a new set of $p$-wave paired FQH states. We also show how some of these new states can be re-expressed in a composite fermion type formulation. In Section V, we review how existing numerical studies relate to our proposed states. In Section VI, we compare our proposed states to existing empirical data. In Section VII, we conclude with a summary of our results and its generalizations, as well as the implications for topological quantum computation.

\section{Hierarchization Generalization}

In the HH picture, hierarchization is carried out by forming a gas of fundamental quasiholes or quasielectrons in an Abelian FQH state that is projected into a Laughlin-type state. We generalize this construction by forming a gas of quasiparticle excitations of specified type in an arbitrary FQH state that is projected into another FQH-type state. The $k^{th}$ level hierarchy wavefunction $\Psi_{k}$, with electron coordinates $z_{j}$ and quasiparticle excitations of arbitrary type (left implicit) at the coordinates $w_{j}$, is obtained from the $\left( k-1 \right)^{th}$ level state $\Psi_{k-1}$ by taking the inner product
\begin{eqnarray}
\label{eq:hierarch}
&& \!\!\!\!\!\!\!\!\!\!\!\!\!\!\!\! \Psi _{k}\left( z_{1},\ldots ,z_{N_{0}};w_{1},\ldots ,w_{n_{k}} \right) \notag \\
&=&\int \prod\limits_{\alpha=1}^{N_{k}}d^{2}u_{\alpha}
\Phi _{k}^{\ast} \left( u_{1},\ldots ,u_{N_{k}};w_{1},\ldots,w_{n_{k}}\right) \notag \\
&&\times \Psi _{k-1}\left( z_{1},\ldots ,z_{N_{0}};u_{1},\ldots ,u_{N_{k}},w_{1},\ldots ,w_{n_{k}}\right)
\end{eqnarray}%
where $u_{j}$ are the coordinates of excitations in the $\left( k-1 \right)^{th}$ level state that form a quasiparticle gas which is projected onto a FQH-type state given by the wavefunction $\Phi _{k}$. These quasiparticles are matched up with the ``electrons'' of the $\Phi _{k}$ state, which must therefore have the same braiding statistics, up to a bosonic factor. Also, the quasiparticle gas excitations (and thus their corresponding ``electrons'') should be Abelian. Together, these ensure single-valuedness of the integrand in the integration coordinates, and thus a well-defined inner product and unique lowest energy $k^{th}$ level ground state ($n_{k}=0$) wavefunction. The specific Abelian excitation type of these quasiparticles, as well as the wavefunctions $\Phi_k$, should be determined by physical arguments, possibly involving energetic considerations and charge minimality.

We note that the concept of quasiparticle ``condensation'' and the formation of these gases in the hierarchy picture should perhaps not be taken too literally (and in fact does not even have a clear meaning in the composite fermion picture~\cite{Jain89}). Nonetheless, it provides an intuitively attractive way of constructing trial wavefunctions for incompressible electron liquids. The introduction of large numbers of quasiholes or quasielectrons into the parent state allows one to lower or raise the filling fraction, but one cannot introduce the quasiparticles in fixed positions, or the new trial wavefunction would not be homogeneous. To regain constant electron density, one should integrate out the coordinates for the quasiparticle gas. This must be done with appropriate weighting for the different quasiparticle configurations, to achieve incompressibility for the new state. Heuristically, one expects that it is a good idea to weight the quasiparticle configurations by the wavefunction for some incompressible state. The hierarchical wavefunction should then satisfy an approximate plasma analogy, if an orthogonality postulate for the parent wavefunction is satisfied~\cite{Read90}. This postulate says that the overlap between electron wavefunctions with different configurations of the quasiparticles vanishes. If this holds, one expects that expectation values for the hierarchy state can be calculated using only a single integral over quasiparticle positions, since the integral over the electron coordinates effectively introduces a delta function into the double integral over the quasiparticle coordinates. This means that the expectation values are approximately those for a classical plasma with two types of particles, corresponding to the electrons and the quasiparticles. Of course the wavefunctions constructed using the hierarchy also satisfy the more basic physical requirements for FQH trial wavefunctions; for example they are antisymmetric and holomorphic in the electron coordinates.

In order for the inner product integrations to give a non-zero result, the number of excitations $N_{k}$ in the quasiparticle gas must be chosen such that the highest power of $u_{\alpha}$ in $\Psi_{k-1}$ is equal to that in $\Phi_k$ (with $u_{\alpha}^{\ast}$ counting as a negative power). This also ensures that the resulting electron wavefunction is  homogeneous. One may think of this as the $\left(k-1\right)^{th}$ level quasiparticle gas determining how many induced ``flux'' quanta are felt by the ``electrons'' in $\Psi_{k-1}$ (where the $0^{th}$ level ``flux'' and ``electrons'' are of course the actual magnetic flux and electrons of the system). This gives a system of equations relating the number of flux quanta $N_{\phi}$, electrons $N_{0}$, quasiparticle gas excitations of each level $N_{j}$ ($j=1,\ldots,k$), and additional quasiparticles. This may be immediately solved to obtain the filling fraction \footnote{In the context of wavefunctions, $\nu$ refers to the filling fraction in the first excited Landau level, rather than the total filling fraction of the described state.} and shift from the expression
\begin{equation}
\label{eq:Nphi}
N_{\phi} = \nu^{-1} N_{0} - S
.
\end{equation}
For clean systems on closed surfaces, the shift $S$ is a topological quantum number that distinguishes different FQH liquids that occur at the same filling fraction~\cite{Wen92c,Wen95}. Clearly, such ideal conditions do not exist in experiments, since physical FQH systems always have some disorder and occur on disklike topologies, so the shift is not experimentally relevant. However, the shift is nonetheless a good indicator in numerical ``experiments,'' where such ideal conditions are achieved more easily than more realistic conditions. When we refer to explicit values of the shift in the following, it should be understood that we specifically mean the shift of the ground state on a sphere, since $S=0$ for ground states on the torus, higher genus surfaces are rarely considered, and excitations can change the shift to essentially any value.

The quasiparticle excitation spectrum of a $k^{th}$ level hierarchy state contains a charge $2e$ boson $B_{0}$ and chargeless bosons $B_{j}$ associated with each level of hierarchization ($j=1,\ldots,k$). These are identified with the vacuum in the anyonic charge spectrum (i.e. quasiparticles that differ only by these bosons have the same anyonic charge), and all permissible quasiparticle excitations must be mutually local (i.e. have trivial monodromy) with them. Allowed excitations must also be mutually local with the electrons or, equivalently, with the charge $e$ fermionic hole $h_{0}$ of the excitation spectrum (two of which combine to give $B_{0}$). We say excitations of $\Psi_{k}$ ``belong to the $j^{th}$ layer'' if they arise entirely from $\Phi_{j}$.

A natural method of generating wavefunctions for FQH states is to use conformal field theory correlators with appropriately chosen vertex operator insertions for the various excitations present~\cite{Moore91}. Excitations from a particular layer can be written as a vertex operator insertion in that layer, but general excitations may involve insertions of operators in multiple layers. To produce the ground state for a hierarchization in which the $j^{th}$ level quasiparticle gas is always formed from excitations belonging only to the $j^{th}$ layer \footnote{It is straightforward to similarly construct hierarchies with gases of multi-layer excitations, but it is cumbersome and unnecessary for the examples in this paper.} we use the CFT correlators%
\begin{eqnarray}
&&\!\!\!\!\!\!\!\! \Phi _{j}\left( u_{1}^{\left(j\right)},\ldots ,u_{N_{j}}^{\left(j\right)};u_{1}^{\left(j+1\right)},\ldots ,u_{N_{j+1}}^{\left(j+1\right)}\right) \notag \\
&&=\left\langle \prod\limits_{\alpha=1}^{N_{j}} V_{e_{j}}\left( u_{\alpha}^{\left(j\right)}\right)
\prod\limits_{\beta=1}^{N_{j+1}} V_{c_{j}}\left( u_{\beta}^{\left(j+1\right)}\right) \right\rangle ,
\label{eq:Phij}
\end{eqnarray}%
where $V_{e_{j}}$ and $V_{c_{j}}$ are respectively vertex operators for the ``electrons'' and quasiparticle gas excitations of the $j^{th}$ layer. Through most of this paper, we employ the standard convention of leaving the neutralizing background charge operators implicit, as well as the Gaussian factors to which they give rise in the resulting wavefunctions. Using this expression in Eq.~(\ref{eq:hierarch}), we take $\Psi _{0} = \Phi _{0}$ with $z = u^{\left(0\right)}$, and $\Phi _{k}$ has no quasiparticle gas and hence no coordinates $u^{\left(k+1\right)}$, nor vertex operators $V_{c_{k}}$.

\section{Reconstructing the Past}
\label{sec:Past}

The particle-hole conjugate~\cite{Girvin84} of an arbitrary FQH state $\psi $ is obtained by projecting holes of a $\nu =1$ quantum Hall wavefunction onto this state. This is carried out by using
\begin{equation}
\label{eq:ph0}
\Psi _{0} =\prod\limits_{\alpha<\beta}\left( z_{\alpha}-z_{\beta}\right)
\prod\limits_{\alpha,\beta}\left( z_{\alpha}-u_{\beta}\right) \prod\limits_{\alpha<\beta}\left( u_{\alpha}-u_{\beta}\right) ,
\end{equation}
\begin{equation}
\Phi _{1} \left( u_{1},\ldots ,u_{N_{1}}\right)=\psi \left( u_{1},\ldots ,u_{N_{1}}\right)
\end{equation}
in Eq.~(\ref{eq:hierarch}). (Index ranges of coordinates will be left implicit from now on.) Eq.~(\ref{eq:ph0}) may be obtained using the vertex operators
\begin{equation}
V_{e_{0}} = V_{c_{0}} = e^{i \varphi_{0} }
\end{equation}
in Eq.~(\ref{eq:Phij}). Solving the system of equations
\begin{eqnarray}
N_{\phi} &=& \left( N_{0} - 1 \right) + N_{1} \\
0 &=& N_{0} + \left( N_{1} -1 \right) - \left( \nu_{\psi}^{-1} N_{1} - S_{\psi} \right)
\end{eqnarray}
imposed by homogeneity of the wavefunction, and comparing to Eq.~\ref{eq:Nphi}, we find the particle-hole conjugate of $\psi$ has
\begin{eqnarray}
\nu &=& 1- \nu_{\psi} \\
S &=& \frac{1-\nu_{\psi}S_{\psi} }{1 - \nu_{\psi}},
\end{eqnarray}
where $\nu_{\psi}$ and $S_{\psi}$ are the filling fraction and shift of $\psi$.

The HH hierarchy~\cite{Haldane83,Halperin84}, combined with particle-hole conjugation, may be used to obtain all the FQH states observed in the first Landau level. This hierarchy is most concisely described as a $\text{U}\left(1\right)_{K}$ CFT or Chern-Simons theory~\cite{Wen92a}, where the coupling-constant $K$-matrix has non-zero elements $K_{00}$ odd, $K_{jj}$ even for $j>0$, and $K_{j,j+1}=K_{j+1,j}=\pm 1$. To make contact between this and explicit wavefunctions using CFT correlators, we use 
\begin{eqnarray}
m_{0} &=& K_{00}, \\
m_{j} &=& K_{jj}-\frac{1}{m_{j-1}} \qquad \text{for } j>0
\end{eqnarray}%
to define the vertex operators
\begin{eqnarray}
\label{eq:HH0}
&& \!\!\!\!\!\!\!\!\!\!
\begin{array}{lll}
V_{e_{0}} = e^{i\sqrt{m_{0}} \, \varphi_{0} }, &
\,\,\,\, V_{\lambda q_{0}} = e^{i\frac{\lambda}{\sqrt{m_{0}}} \varphi_{0} } & \\
\end{array} \\
\label{eq:HHj}
&& \!\!\!\!\!\!\!\!\!\!\!\!\!\!\! \left\{
\begin{array}{lll}
V_{e_{j}} = e^{i\sqrt{m_{j}} \, \overline{\varphi} _{j} }, &
V_{\lambda q_{j}} = e^{i\frac{\lambda}{\sqrt{m_{j}}}\overline{\varphi} _{j} } &
\text{for  } m_{j}>0 \\
V_{e_{j}}= e^{i\sqrt{-m_{j}} \, \varphi _{j} }, &
V_{\lambda q_{j}} = e^{-i\frac{\lambda}{\sqrt{-m_{j}}}\varphi_{j} } &
\text{for  } m_{j}<0
\end{array}%
\right. \\
\label{eq:HHc}
&& \!\!\!\!\!\!\!\!\!\!
\begin{array}{lll}
V_{c_{j}} = V_{K_{j,j+1} q_{j}} & & \\
\end{array}
\end{eqnarray}%
used in Eq.~(\ref{eq:Phij}) to give
\begin{widetext}
\begin{eqnarray}
&& \!\!\!\!\!\!\!\!\!\!
\begin{array}{ll}
\Psi_{0} = \prod\limits_{\alpha<\beta}\left( z_{\alpha}-z_{\beta}\right)^{m_{0}} \prod\limits_{\alpha,\beta}\left( z_{\alpha}-u_{\beta}^{\left(1\right)}\right)^{K_{01}} \prod\limits_{\alpha<\beta}\left( u_{\alpha}^{\left(1\right)}-u_{\beta}^{\left(1\right)}\right)^{\frac{1}{m_{0}}}
& \\
\end{array} \\
&& \!\!\!\!\!\!\!\!\!\!\!\!\!\!\! \left\{
\begin{array}{ll}
\Phi_{j} = \prod\limits_{\alpha<\beta}\left( u_{\alpha}^{ \left(j\right) \ast}-u_{\beta}^{ \left(j\right) \ast}\right)^{m_{j}} \prod\limits_{\alpha,\beta}\left( u_{\alpha}^{\left(j\right) \ast}-u_{\beta}^{\left(j+1\right) \ast}\right)^{K_{j,j+1}} \prod\limits_{\alpha<\beta}\left( u_{\alpha}^{\left(j+1\right) \ast}-u_{\beta}^{\left(j+1\right) \ast}\right)^{\frac{1}{m_{j}}} &
\qquad \text{for  } m_{j}>0 \\
\Phi_{j} = \prod\limits_{\alpha<\beta}\left( u_{\alpha}^{\left(j\right)}-u_{\beta}^{\left(j\right)}\right)^{-m_{j}} \prod\limits_{\alpha,\beta}\left( u_{\alpha}^{ \left(j\right)}-u_{\beta}^{ \left(j+1\right)}\right)^{-K_{j,j+1}} \prod\limits_{\alpha<\beta}\left( u_{\alpha}^{ \left(j+1\right)}-u_{\beta}^{ \left(j+1\right)}\right)^{-\frac{1}{m_{j}}}
&
\qquad \text{for  } m_{j}<0
\end{array}%
\right.
.
\end{eqnarray}%
\end{widetext}
Substituting these into Eq.~(\ref{eq:hierarch}) reproduces the hierarchy wavefunctions given in Ref.~\onlinecite{Read90}, if we allow a negative sign in the exponent to be treated as equivalent to complex-conjugation. To reproduce the hierarchy wavefunctions given in Ref.~\onlinecite{Blok91}, one must replace the terms with negative exponents generated by vertex operators with negative $\lambda$ (e.g. those with $K_{j,j+1}=-\text{sgn}\left\{m_{j}\right\}$) by terms with matching positive powers of the complex-conjugated variables. Furthermore, a projection of the wavefunction into the lowest Landau level needs to be applied at the end (replacing $z^{\ast}$ with $2 \partial /\partial z$) when $K_{01}=-1$. Exchanging negative powers for complex conjugates followed by lowest Landau level projection should only introduce short-ranged effects and will leave the universal properties of the states unaffected, as discussed in Ref.~\onlinecite{Read90}.

For states constructed using quasielectrons, we have used Laughlin's ansatz~\cite{Laughlin83} for the quasielectrons' wavefunctions in the formulas above. Alternatively, one may use Jain's ansatz~\cite{Jain89}. This may in fact be advantageous, since numerical studies indicate that the latter has a better behaved charge density and statistics parameter for systems of approximately $100$ electrons~\cite{Kjonsberg99a,Kjonsberg99b}. It is straightforward to write explicit wavefunctions with this alternate description of quasielectrons and use them in the above hierarchy. Though this will not exactly match the previously proposed hierarchy wavefunctions, the universal properties should be the same.

The quasiparticle gas excitations $c_{j} = \pm q_{j}$ are either the fundamental quasiholes or quasielectrons of the $j^{th}$ level state $\Psi_{j}$, the choice of which should be determined by whether the filling fraction is respectively decreased or increased in going to the next level. Hence, one ought to use
\begin{equation}
K_{j,j+1}=-\text{sgn}\left\{K_{j+1,j+1}\right\}.
\end{equation}
The filling fraction and shift may be determined from $K$ directly to be%
\begin{eqnarray}
\label{eq:HHnu}
\nu &=&\left[ K^{-1}\right] _{00} = \frac{1}{K_{00} - \frac{1}{ K_{11} - \frac{1}{\ldots - \frac{1}{K_{kk}}}}}, \\
\label{eq:HHS}
S &=& \frac{1}{\nu }\sum_{j=0}^{k}\left[ K^{-1}\right] _{0j}K_{jj}
\end{eqnarray}%
by solving the system of equations~\cite{Wen92a}
\begin{equation}
\delta_{0i} N_{\phi} = \sum_{j=0}^{k} K_{ij} \left( N_{j} - \delta_{ij} \right)
.
\end{equation}

An arbitrary HH quasiparticle excitation is specified by the number of ``fluxes'' (vortices) $a_{j} \in \mathbb{Z}$ in the $j^{th}$ layer. An $\overrightarrow{a}$ excitation produces a factor of
\begin{equation}
\prod_{j=0}^{k} \prod_{\alpha_{j}} \left(w-u_{\alpha_{j}}^{\left(j\right)}\right)^{a_{j}}
\end{equation}
in the wavefunction $\Psi_{k}$, and is obtained by inserting
\begin{equation}
\label{eq:HHa}
V_{\overrightarrow{a}} \left(w\right) = \prod\limits_{j=0}^{k} V_{\lambda_{j}q_{j}}\left(w\right),
\end{equation}%
where $\lambda_{0}=a_{0}$ and $\lambda_{j>0} = a_{j} - \frac{K_{j,j-1} \lambda_{j-1}}{m_{j-1}}$, in the CFT correlator. The electric charges and braiding statistics (in terms of $R$-symbols) of such excitations are given by%
\begin{eqnarray}
\label{eq:HHQ}
Q_{\overrightarrow{a}} &=&e \,\, \hat{t}_{0}\cdot K^{-1}\cdot \overrightarrow{a} \\
\label{eq:HHR}
R^{\overrightarrow{a},\overrightarrow{b}} &=&\exp \left( i\pi \,
\overrightarrow{a}\cdot K^{-1}\cdot \overrightarrow{b}\right)
,
\end{eqnarray}%
where $\hat{t}_{j}$ is the unit vector with a $1$ in the $j^{th}$ row. In the HH hierarchy states, we have the anyonic charges for holes $h_{0} = K\cdot \hat{t}_{0}$, and bosons $B_{0} = 2h_{0}$ and $B_{j>0} = K\cdot \hat{t}_{j}$. Using the appropriate identifications, the entire excitation spectrum in this case is generated, through repeated fusion, by the fundamental quasihole excitation in the highest hierarchy layer, so arbitrary excitations may be written as $n\hat{t}_{k}$. These states have $\left| \det K \right|$-fold ground state degeneracy on the torus.

There is an alternate description of the FQH states observed at $\nu = \frac{n}{ 2pn \pm 1}$ in terms of composite fermions~\cite{Jain89} given by the wavefunctions
\begin{equation}
\label{eq:CF}
\Psi_{\nu}^{\left( \text{CF} \right)} = \mathcal{P}_{LLL} \left\{ \chi_{1}^{2p} \chi_{\pm n} \right\}
\end{equation}
where $\mathcal{P}_{LLL}$ is the projection onto the lowest Landau level and $\chi_{n}$ is the wavefunction for $n$ filled Landau levels. It is well known~\cite{Read90} that these composite fermion wavefunctions have the same universal properties (i.e. filling fraction, quasiparticle charge, and braiding statistics) as the $k^{th}$ level HH hierarchy states for $k=n-1$ with
\begin{equation}
K_{00} = 2p+1 , \qquad K_{jj}=2 \quad \text{for }j>0
\end{equation}
for $\nu = \frac{n}{ 2pn + 1}$; and with
\begin{equation}
K_{00} = 2p-1 , \qquad K_{jj}=-2 \quad \text{for }j>0
\end{equation}
for $\nu = \frac{n}{ 2pn - 1}$. The HH and composite fermion descriptions can also be shown to produce the same shift $S=2p \pm n$ for these states. Despite the fact that these are just different descriptions of the same topological orders~\cite{Wen95} (i.e.~universality classes), the composite fermion approach produces different explicit wavefunctions that have some advantages, particularly regarding their employment in numerical studies and the description of quasielectrons (for a thorough discussion of these and other issues, see e.g. Ref.~\onlinecite{Jain07} and references therein). On the other hand, the composite fermion approach has a shortcoming in that it must still incorporate some sort of hierarchy. In particular, there are many experimentally observed FQH states that the $\nu = \frac{n}{ 2pn \pm 1}$ series can only account for by using particle-hole conjugation, which (as we have seen) is just a special case of a hierarchical construction; and, even with particle-hole conjugation, this series cannot account for some experimentally observed lowest Landau level FQH states, such as those at $\nu=4/11$, $5/13$, etc.~\cite{Pan03} In any case, the topological order of a state is much more robust than the particular details of any of its trial wavefunctions, and it is the universal properties that can and will distinguish states experimentally. Since our primary interest in this paper is in the identification of new states and their topological order, we will dwell on such issues no longer, and simply state that in circumstances where it could be advantageous, one should consider employing the composite fermion picture.

\section{Building On Moore--Read}

The general hierarchy prescription in Eq.~(\ref{eq:hierarch}) can generate a multitude of states at any given filling fraction, so we will restrict our attention to the constructions that seem most physically relevant and tenable. Specifically, we build the simplest possible hierarchy involving the MR state, which is closely analogous to the HH hierarchy in that the hierarchization occurs only in the $\text{U}(1)$ charge sector of the theory.
This is perhaps the most natural way to form hierarchies with non-Abelian states in general, because it treats the mechanism giving rise to the non-Abelian sector (in the MR case: pairing giving rise to the Ising sector) as a ubiquitous property of the class of states, while the charge sector is allowed to form a hierarchy as it is already known to do for Abelian states. We will also construct a somewhat less simple hierarchy over MR which involves the Ising sector in a non-trivial way, but for which the additional layers are all $\text{U}(1)$s.

We begin with the MR state at the $0^{th}$ level of hierarchy. The CFT describing MR may be written as\footnote{Normally the Abelian charge sector for the MR state would be written as $\text{U}\left( 1\right) _{4}$, but we can instead write it as $\text{U}\left( 1\right) _{2}$ by allowing the spectrum to include half-integer fluxes.}
\begin{eqnarray}
&&\left. \text{Ising}\times \text{U}\left( 1\right) _{2}\right| _{\mathcal{C}}, \\
&&\mathcal{C}=\left\{ \left( I,n \right), \left( \psi,n \right), \left( \sigma,n+1/2 \right): n \in \mathbb{Z} \right\}
.
\end{eqnarray}
The Ising sector's anyonic charges (CFT primary fields): $I$ (vacuum), $\psi$ (Majorana fermion), and $\sigma$ (spin) obey the commutative fusion algebra
\begin{equation}
\begin{array}{lll}
I \times I = I, \quad & I \times \psi = \psi, \quad & I \times \sigma = \sigma, \\
\psi \times \psi = I, \quad & \psi \times \sigma = \sigma, \quad & \sigma \times \sigma = I+\psi .
\end{array}
\label{eq:Isingfusion}
\end{equation}
The entire anyonic charge spectrum $\mathcal{C}$ is generated by the fundamental quasihole, $\left(\sigma,1/2\right)$. The corresponding electron and fundamental quasihole vertex operators are respectively%
\begin{equation}
\label{eq:BS0}
V_{e_{0}} =\psi  e^{i\sqrt{2} \, \varphi_{0} }, \quad
V_{qh_{0}} =\sigma e^{i\frac{1}{\sqrt{8}} \varphi_{0} }.
\end{equation}%

To form a hierarchy over the MR state, we must first specify the $0^{th}$ level quasiparticle gas. The physical picture we envision here is that forming a gas of fundamental quasiholes/quasielectrons
$\left(\sigma,\pm 1/2\right)$ of the MR state forces them to pair up into preferential Abelian bound state excitations that can no longer be recoupled. A pair of $\sigma$ Ising charges have two possible fusion channels, $I$ and $\psi$, that describe their combined anyonic charge. If the quasiparticles are well separated, the two channels are degenerate, but because of the density of quasiparticles required for the hierarchy construction, they will necessarily be in close proximity and so this degeneracy will be broken. It is not a priori clear which fusion channel is favored and in fact this may be difficult to determine from first principles, as there could be many-quasiparticle effects. However, we expect that the favored channel will be the one that leads to a fusion product that is just a Laughlin-type quasiparticle of charge $\pm e/2$. This will lead to the simplest hierarchy wavefunctions, which most resemble the HH hierarchy. For the MR state, this means favoring the $I$ channel, so we expect the $0^{th}$ quasiparticle gas to be composed of excitations with anyonic charge $\left(I,K_{01}\right)$, where $K_{01}=\pm 1$ indicates paired quasiholes/quasielectrons. The corresponding vertex operator and resulting wavefunction are%
\begin{eqnarray}
\label{eq:VcMR}
V_{c_{0}} &=& I e^{i\frac{K_{01}}{\sqrt{2}}\varphi_{0} }, \\
\Psi _{0}&=&\text{Pf}\left\{ \frac{1}{z_{\alpha}-z_{\beta}}\right\} \prod\limits_{\alpha<\beta}\left(
z_{\alpha}-z_{\beta}\right) ^{2} \notag \\
&& \quad \quad \times \prod\limits_{\alpha,\beta}\left( z_{\alpha}-u_{\beta}\right)^{K_{01}}
\prod\limits_{\alpha<\beta}\left( u_{\alpha}-u_{\beta}\right) ^{1/2}
\label{eq:Psi0MR}
,
\end{eqnarray}
where $\text{Pf}$ is the Pfaffian, defined for a $2n \times 2n$ antisymmetric matrix $A_{\alpha \beta}$ by
\begin{equation}
\text{Pf}\left\{ A_{\alpha \beta} \right\} = \frac{1}{\left(2n\right)!!} \sum_{\sigma \in S_{2n}} \text{sgn}\left(\sigma\right) \prod_{j= 1}^{n} A_{\sigma \left(2j-1\right) \sigma\left(2j\right)}
.
\end{equation}

In order to build the simplest hierarchy over MR, we take all higher layers to be Abelian $\text{U} \left( 1\right)$ Hall fluids, with the minimal charge excitations of each level comprising its quasiparticle gas. It follows that each level's quasiparticle gas excitations are trivial in the Ising sector, and hence we may again use the $K$-matrix formalism to describe the resulting hierarchy states as
\begin{eqnarray}
&&\left. \text{Ising}\times \text{U} \left( 1\right) _{K}\right| _{\mathcal{C}}, \\
&&\mathcal{C} = \left\{
\begin{array}{lcl}
\left( I, \overrightarrow{a}\right) & : & a_{j} \in \mathbb{Z}, \\
\left( \psi, \overrightarrow{a}\right) & : & a_{j} \in \mathbb{Z}, \\
\left( \sigma, \overrightarrow{a}\right) & : & a_{0} \in \mathbb{Z} +\frac{1}{2}, \,\, a_{j>0} \in \mathbb{Z}
\end{array}
\right\}
\end{eqnarray}
where now $K_{00}=2$ (rather than $K_{00}$ odd) and, as before, the other non-zero elements of $K$ are $K_{jj}$ even for $j>0$ and $K_{j-1,j}=K_{j,j-1}=- \text{sgn}\left\{ K_{jj} \right\}$. The anyonic charges in the spectrum $\mathcal{C}$ are given by
$A=\left( a_{\text{I}},\overrightarrow{a}\right) $ where
$a_{\text{I}}$ is the Ising charge and $\overrightarrow{a}$
is the U$\left( 1\right) _{K}$ flux vector. We obtain explicit $\Phi_{j>0}$ for use with Eqs.~(\ref{eq:hierarch},\ref{eq:Psi0MR}), by simply applying Eqs.~(\ref{eq:Phij},\ref{eq:HHj},\ref{eq:HHc}) for the new $K$. The filling fraction and shift are
\begin{eqnarray}
\nu &=& \nu_{K} \\
S &=& S_{K} +1 ,
\end{eqnarray}
where $\nu_{K}$ and $S_{K}$ are given in Eqs.~(\ref{eq:HHnu},\ref{eq:HHS}) respectively, and the $+1$ is due to the Pfaffian from the Ising sector.

An arbitrary $A = \left( a_{\text{I}},\overrightarrow{a}\right) $ quasiparticle excitation corresponds to insertion of the vertex operator
\begin{equation}
V_{A} \left(w \right) = a_{\text{I}}\left(w \right) V_{\overrightarrow{a}}\left(w \right)
\end{equation}
with $V_{\overrightarrow{a}}$ from Eq.~(\ref{eq:HHa}), but now permitting half-integral $a_{0}$. These quasiparticle excitations have the electric charge and the braiding $R$-symbols given by
\begin{eqnarray}
Q_{A} &=& Q_{\overrightarrow{a}} \\
R_{C}^{A,B} &=& R_{c_{\text{I}}}^{a_{\text{I}},b_{\text{I}}}R^{\overrightarrow{a},\overrightarrow{b}}
,
\end{eqnarray}
where $Q_{\overrightarrow{a}}$ and $R^{\overrightarrow{a},\overrightarrow{b}}$ are given in Eqs.~(\ref{eq:HHQ},\ref{eq:HHR}) respectively, and
the Ising sector's $R_{c_{\text{I}}}^{a_{\text{I}},b_{\text{I}}}$ are%
\begin{eqnarray}
&& \!\!\!\!\!\!\!\! R_{I}^{I,I}=R_{\psi}^{I,\psi}=R_{\psi}^{\psi,I}=R_{\sigma}^{I,\sigma}=R_{\sigma}^{\sigma,I}=1,
\quad R_{I}^{\psi ,\psi }=-1, \notag \\
&& \!\!\!\!\!\!\!\! R_{\sigma}^{\psi ,\sigma }=R_{\sigma }^{\sigma ,\psi }=-i, \quad R_{I}^{\sigma ,\sigma}=e^{-i\frac{\pi }{8}},\quad R_{\psi }^{\sigma ,\sigma }=e^{i\frac{3\pi }{8}}
.
\label{eq:Isingbraiding}
\end{eqnarray}%
In these states, we have $h_{0} = \left( \psi ,K\cdot \hat{t}_{0}\right)$, $B_{0} = \left( I,2K\cdot \hat{t}_{0}\right)$, and $B_{j>0} = \left( I,K\cdot \hat{t}_{j}\right)$. Two quasiparticle excitation types are needed to generate the entire charge spectrum: the fundamental quasiholes/quasielectrons in the lowest and highest layers, $qh_{0} = \left( \sigma ,\frac{1}{2}\hat{t}_{0}\right)$ and
$q_{k} = \left( I,\hat{t}_{k}\right)$. The resulting states have $3 \left|\det K \right|$-fold ground state degeneracy on the torus when the system has an even number of electrons $N_{0}$, and $\left|\det K \right|$-fold torus degeneracy when $N_{0}$ is odd.

We obtain a $\nu =2/3$ ground state wavefunction at the $1^{st}$ level of hierarchy by using
$K=\left[
\begin{array}{rr}
2 & -1 \\
-1 & 2
\end{array}
\right]$, for which%
\begin{equation}
\label{eq:twothirds}
\Phi _{1}\left( u_{1},\ldots ,u_{N_{1}}\right)  =\prod\limits_{\alpha<\beta}\left( u_{\alpha}^{\ast}-u_{\beta}^{\ast}\right) ^{3/2}
.
\end{equation}
This state has $S=4$, a $9$-fold ground state degeneracy on the torus (for $N_{0}$ even), and the spectrum generating excitations $qh_{0}$ and $q_{1}$ have minimal electric charge $e/3$. This provides a candidate state for $\nu =8/3$.

Alternatively, we obtain a $\nu =2/5$ ground state wavefunction at the $1^{st}$ hierarchy level by using
$K=\left[
\begin{array}{rr}
2 & 1 \\
1 & -2
\end{array}
\right]$, for which%
\begin{equation}
\label{eq:twofifths}
\Phi _{1}\left( u_{1},\ldots ,u_{N_{1}}\right)  =\prod\limits_{\alpha<\beta}\left( u_{\alpha}-u_{\beta}\right) ^{5/2}
.
\end{equation}%
This state has $S=2$, a $15$-fold ground state degeneracy on the torus (for $N_{0}$ even), and the spectrum generating excitations $qh_{0}$ and $q_{1}$ have minimal electric charge $e/5$. This provides a candidate state for $\nu=12/5$.

To be completely explicit, we plug Eqs.~(\ref{eq:Psi0MR},\ref{eq:twothirds},\ref{eq:twofifths}) into Eq.~(\ref{eq:hierarch}), follow the prescription for dealing with negative exponents as discussed for the Abelian hierarchy in Section~\ref{sec:Past}, and include the appropriate Gaussian factors to obtain the $\nu = 2/3$ and $2/5$ wavefunctions
\begin{widetext}
\begin{eqnarray}
\label{eq:extwothirds}
\Psi_{2/3}^{\left(\text{BS}\right)} &=&\int \prod \limits_{\mu = 1}^{N_{1}} d^{2}u_{\mu} e^{-\frac{1}{4 \ell^{2}} \sum\limits_{\alpha =1}^{N_{0}} \left| z_{\alpha} \right|^{2} -\frac{1}{4 \ell^{2}} \sum\limits_{\alpha =1}^{N_{1}} \left| u_{\alpha} \right|^{2} } \prod\limits_{\alpha < \beta} \left( u_{\alpha}-u_{\beta} \right)^{2}  \prod\limits_{\alpha,\beta} \left( u_{\alpha}^{\ast}- 2\frac{\partial}{\partial z_{\beta}} \right) \text{Pf} \left\{ \frac{1}{z_{\alpha}-z_{\beta}} \right\} \prod\limits_{\alpha < \beta} \left( z_{\alpha}-z_{\beta} \right)^{2} \\
\label{eq:extwofifths}
\Psi_{2/5}^{\left(\text{BS}\right)} &=& \int \prod \limits_{\mu = 1}^{N_{1}} d^{2}u_{\mu} e^{-\frac{1}{4 \ell^{2}} \sum\limits_{\alpha =1}^{N_{0}} \left| z_{\alpha} \right|^{2} -\frac{1}{4 \ell^{2}} \sum\limits_{\alpha =1}^{N_{1}} \left| u_{\alpha} \right|^{2} } \prod\limits_{\alpha < \beta} \left( u_{\alpha}^{\ast}-u_{\beta}^{\ast} \right)^{2} \left|  u_{\alpha}-u_{\beta} \right|  \prod\limits_{\alpha,\beta} \left( u_{\alpha}-z_{\beta} \right) \text{Pf} \left\{ \frac{1}{z_{\alpha}-z_{\beta}} \right\} \prod\limits_{\alpha < \beta} \left( z_{\alpha}-z_{\beta} \right)^{2} \qquad
\end{eqnarray}
\end{widetext}
where $N_{1} = \frac{1}{2} N_{0} +1$ and $\ell=\sqrt{ \frac{ \hbar c}{ e B} }$ is the magnetic length. We remind the reader that the previous discussion regarding alternate descriptions of quasielectrons should be kept in mind for the wavefunction in Eq.~(\ref{eq:extwothirds}).

We can obtain a $\nu = 3/8$ state at the $2^{nd}$ level of hierarchy by using $K_{11}=K_{22}=-2$ [i.e. building on the $\nu=2/5$ state from Eq.~(\ref{eq:extwofifths})]. This gives a candidate state to describe what may be a FQH state developing at $\nu = 19/8$ seen in Refs.~\onlinecite{Xia04,Pan08}. From this we can predict that the next state to appear in the second Landau level after $\nu = 19/8$ would be the next level of hierarchy occurring at $\nu = 26/11$ (i.e. $\nu = 2 + 4/11$ obtained from $K_{11}=K_{22}=K_{33}=-2$).

We also note that a $\nu = 4/5$ state is produced at the $3^{rd}$ level of hierarchy by using $K_{11}=K_{22}=K_{33}=2$, but, as it must pass through an unobserved $\nu = 3/4$ state ($\nu = 9/4$ and $11/4$ appear to be Fermi liquids of composite fermions~\cite{Pan08}) in the $2^{nd}$ hierarchy level, this is rather unlikely to be the correct description for the observed $\nu = 14/5$ plateau, which is expected to be a Laughlin state anyway.

If the MR quasiholes/quasielectrons were instead to pair up in the $\psi$-channel to form a hierarchy's $0^{th}$ level quasiparticle gas of $\left(\psi,K_{01}\right)$ excitations, we would have%
\begin{eqnarray}
V_{c_{0}} &=&\psi e^{i\frac{K_{01}}{\sqrt{2}}\varphi_{0} }, \\
\Psi _{0}&=&\text{Pf}\left\{ \frac{1}{Z_{\alpha}-Z_{\beta}}\right\} \prod\limits_{\alpha<\beta}\left(
z_{\alpha}-z_{\beta}\right) ^{2} \notag \\
&& \times \prod\limits_{\alpha,\beta}\left( z_{\alpha}-u_{\beta}^{\left(1\right)}\right)^{K_{01}}
\prod\limits_{\alpha<\beta}\left( u_{\alpha}^{\left(1\right)}-u_{\beta}^{\left(1\right)}\right) ^{1/2}
\end{eqnarray}
instead of Eqs.~(\ref{eq:VcMR},\ref{eq:Psi0MR}), where $Z_{\alpha} = z_{\alpha}$ for $\alpha=1,\ldots,N_{0}$ and $Z_{\alpha + N_{0} } = u_{\alpha}$ for $\alpha=1,\ldots,N_{1}$. Taking all higher layers to be Abelian again gives a hierarchy described by $\left. \text{Ising}\times \text{U} \left( 1\right) _{K}\right| _{\mathcal{C}}$ and Eqs.~(\ref{eq:HHj},\ref{eq:HHc}), but now with $K_{11}$ odd in order to match the braiding statistic of the $c_{0}$ excitations. In this case, the first layer's chargeless boson is $B_{1} = \left( \psi,K\cdot \hat{t}_{1}\right)$ and excitations with $a_{\text{I}} = \sigma$ must have $a_{0},a_{1} \in \mathbb{Z}+\frac{1}{2}$ (and hence cannot be written as a single layer excitation). Specifically, the charge spectrum is given by
\begin{equation}
\mathcal{C} = \left\{
\begin{array}{lcl}
\left( I, \overrightarrow{a}\right) & : & a_{j} \in \mathbb{Z}, \\
\left( \psi, \overrightarrow{a}\right) & : & a_{j} \in \mathbb{Z}, \\
\left( \sigma, \overrightarrow{a}\right) & : & a_{0},a_{1} \in \mathbb{Z} +\frac{1}{2}, \,\, a_{j>1} \in \mathbb{Z}
\end{array}
\right\}
.
\end{equation}
For these states, the filling fraction and shift are
\begin{eqnarray}
\nu &=& \nu_{K} \\
S &=& S_{K} + 1 + \nu^{-1} \left[K^{-1}\right]_{01} ,
\end{eqnarray}
where $\nu_{K}$ and $S_{K}$ are again given by Eqs.~(\ref{eq:HHnu},\ref{eq:HHS}) respectively.

We obtain a $\nu=1/3$ ground state wavefunction at $1^{st}$ level of hierarchy using
$K=\left[
\begin{array}{rr}
2 & 1 \\
1 & -1
\end{array}
\right]$, for which%
\begin{equation}
\label{eq:onethirds}
\Phi _{1}\left( u_{1},\ldots ,u_{N_{1}}\right)
= \prod\limits_{\alpha<\beta}\left( u_{\alpha}-u_{\beta}\right) ^{3/2}
.
\end{equation}%
This state has $S=3$, a $9$-fold ground state degeneracy on the torus (for $N_{0}$ even), and the spectrum is generated by two minimal electric charge $e/3$ excitations, $\left( \sigma,1/2,1/2 \right)$ and $\left( I,0,1 \right)$. This provides a candidate state for $\nu = 7/3$.

We obtain a $\nu=3/5$ ground state wavefunction at $1^{st}$ level of hierarchy using
$K=\left[
\begin{array}{rr}
2 & -1 \\
-1 & 3
\end{array}
\right]$, for which%
\begin{equation}
\label{eq:threefifths}
\Phi _{1}\left( u_{1},\ldots ,u_{N_{1}}\right)
= \prod\limits_{\alpha<\beta}\left( u_{\alpha}^{\ast}-u_{\beta}^{\ast}\right) ^{5/2}
.
\end{equation}%
This state has $S=13/3$, a $15$-fold ground state degeneracy on the torus (for $N_{0}$ even), and the spectrum is generated by two minimal electric charge $e/5$ excitations, $\left( \sigma,1/2,1/2 \right)$ and $\left( I,0,1 \right)$.

One can also construct these kinds of hierarchies over the particle-hole conjugate of the MR state, which recent experiments seem to indicate may in fact describe the $\nu=5/2$ plateau~\cite{Radu08}. This is exactly the same as taking the particle-hole conjugate of the hierarchy states we have built on the MR state. In particular, particle-hole conjugating the state in Eq.~(\ref{eq:extwothirds}) gives a candidate for $\nu = 7/3$ (with $S=-5$); particle-hole conjugating the state constructed from Eq.~(\ref{eq:onethirds}) gives a candidate for $\nu = 8/3$ (with $S=0$); and particle-hole conjugating the state constructed from Eq.~(\ref{eq:threefifths}) gives a candidate for $\nu = 12/5$ (with $S=-4$).

As it may be advantageous for numerical studies, it is expedient to reformulate these hierarchical states using a composite fermion type construction of wavefunctions. This is straightforward for the hierarchies in which the $0^{th}$ level quasiparticles pair up in the trivial channel of the non-Abelian sector, because the hierarchization is entirely in the $\text{U}\left(1\right)$ charge sector, and thus one simply applies a transformation similar to the one that relates the HH hierarchy to the composite fermion description of the Abelian states. In particular, the $\nu = \frac{n}{\left(2p+1\right)n \pm 1}$ states with ground state wavefunctions
\begin{equation}
\label{eq:BSCF}
\Psi_{\nu}^{\left(\text{BS-CF} \right)} = \mathcal{P}_{LLL} \left\{ \text{Pf} \left\{ \frac{1}{z_{\alpha}-z_{\beta}} \right\} \chi_{1}^{2p+1} \chi_{\pm n} \right\}
\end{equation}
(which have $S=2p + 2 \pm n$) are equivalent to the $k^{th}$ level $\text{Ising} \times \text{U}\left(1\right)_{K}$ hierarchy states for $k=n-1$ with
\begin{equation}
K_{00} = 2p+2 , \qquad K_{jj}=2 \quad \text{for }j>0
\end{equation}
for $\nu = \frac{n}{\left(2p+1\right)n + 1}$; and with
\begin{equation}
K_{00} = 2p , \qquad K_{jj}=-2 \quad \text{for }j>0
\end{equation}
for $\nu = \frac{n}{\left(2p+1\right)n - 1}$. These wavefunctions can be compactly written in the suggestive form
\begin{equation}
\label{eq:BSCFapprox}
\Psi_{\nu}^{\left(\text{BS-CF} \right)} = \Psi^{\left(\text{MR} \right)}_{1} \Psi_{\frac{n}{2pn \pm 1}}^{\left( \text{CF} \right)}
\end{equation}
where $\Psi^{\left(\text{MR} \right)}_{1}$ is a $\nu=1$ bosonic MR wavefunction and $\Psi_{\frac{n}{2pn \pm 1}}^{\left( \text{CF} \right)}$ is the composite fermion wavefunction from Eq.~(\ref{eq:CF}). Clearly, these are not exactly the same as the wavefunctions in Eq.~(\ref{eq:BSCF}), since the lowest Landau level projection no longer acts on their entirety, but one might expect it to be a reasonably good approximation. In any case, the choice of partitioning for lowest Landau level projection does not affect the universal properties, so the difference between these is only relevant for numerical studies.

Since we want states built on the $\nu = 5/2$ MR state, we focus on the two $K_{00}=2$ series
\begin{eqnarray}
\label{eq:BSCF1}
\Psi_{\frac{n}{n+1}}^{\left( \text{BS-CF} \right)} &=& \mathcal{P}_{LLL} \left\{ \text{Pf} \left\{ \frac{1}{z_{\alpha}-z_{\beta}} \right\} \chi_{1} \chi_{n} \right\} ,\\
\label{eq:BSCF3}
\Psi_{\frac{n}{3n-1}}^{\left( \text{BS-CF} \right)} &=& \mathcal{P}_{LLL} \left\{ \text{Pf} \left\{ \frac{1}{z_{\alpha}-z_{\beta}} \right\} \chi_{1}^{3} \chi_{-n} \right\}
,
\end{eqnarray}
which have shifts $S=2+n$ and $4-n$, respectively. For $n=2$, these wavefunctions respectively give alternative descriptions of the $\nu = 2/3$ and $2/5$ ground states constructed in Eqs.~(\ref{eq:extwothirds},\ref{eq:extwofifths}).

\section{Comparison to Numerical Studies}

In the absence of sufficient definitive experimental evidence, FQH theorists have traditionally turned to numerical studies as another tool to help educate our guesses as to the likelihood of various proposals' correctness. As it has historically been difficult to exclude candidate states from contention using the available experimental data, a number of quantities that are experimentally unobservable but suitable for numerical evaluation have risen to prominence as indicators of the quality of trial wavefunctions. These include the shift, the ground state degeneracy on the torus, and the overlap for small systems (usually on a sphere) with wavefunctions calculated by exact diagonalization of various Hamiltonians for lowest Landau level electrons. Clearly, it is of great interest to perform such numerical analyses also for our trial wavefunctions. This should be possible using the methods of Refs.~\onlinecite{Jain97,Park98} with the composite fermion formulation of Eqs.~(\ref{eq:BSCF1},\ref{eq:BSCF3}), perhaps with some careful thought regarding the quality of approximation obtained when moving certain terms outside the lowest Landau level projections. We plan to address such numerical matters more thoroughly in a future publication~\cite{BondersonWIP}. Here we will give a brief discussion of previous numerical work on second Landau level states.

Unfortunately, it seems that there is very little numerical work on the $\nu=7/3$ and $8/3$ FQH plateaus, apart from Ref.~\onlinecite{Wojs01}, in which the electron-electron correlation functions seem to favor non-Abelian states at these filling fractions.

For $\nu=12/5$, a study of zero-thickness systems on a sphere with $N_{0}=15$ and $18$ was carried out in Ref.~\onlinecite{Read99}. This work shows that, as the pseudopotentials of the model Hamiltonian used for the exact diagonalization are varied around the Coulomb point, there is a region where the $k=3$ RR state has good overlap with the numerically obtained wavefunction, and a region where the Abelian composite fermion wavefunction has good overlap with the numerically obtained wavefunction. However, there is also a region between these where the overlap for the RR state drops to zero and where the overlap for the Abelian wavefunction is still very small. From this, one can draw the conclusion that there is room for another phase at $\nu=12/5$. A flux scan for $8 \leq N_{0} \leq 14$ with second Landau level Coulomb interactions on the sphere using the techniques of Ref.~\onlinecite{Feiguin08a} reveals~\cite{Feiguin-private} a $\nu =12/5$ state with $S=2$. This matches our $\nu=12/5$ state given in Eq.~(\ref{eq:extwofifths}), but neither the Abelian hierarchy nor the RR states, which respectively have $S=4$ and $S=-2$.

A numerical study including non-zero thickness for systems on a torus with $N_{0}=15$ and $18$ was carried out in Ref.~\onlinecite{Rezayi06}. The ground state degeneracy for the Coulomb potential is observed to be $5$-fold and $10$-fold for different regions in the parameter space. For $N_{0}=15$ (i.e. $N_{0}$ odd), the observed $5$-fold torus ground state degeneracy is consistent with our $\nu=2/5$ states, though it is also consistent with an Abelian hierarchy state. For $N_{0}=18$, a $15$-fold ground state degeneracy consistent with our state did not appear. The $10$-fold ground state degeneracy (for both $N_{0}$) is consistent with the $k=3$ RR state, however, the splitting of these ground states is not as small for $N_{0}=18$ as it is for $N_{0}=15$, and also the gap (from these to the next states) is smaller, even comparable to the ground state splitting. A $15$-fold torus ground state degeneracy may well emerge when physical effects, such as Landau level mixing and particle-hole symmetry-breaking are incorporated in the simulations, or simply if larger system sizes or a larger region of the pseudopotential and layer thickness parameter space are explored. Furthermore, numerical studies for systems on a disk with boundary edge have found that edge effects can play a significant role in determining which phase is stabilized~\cite{Wan08a}, so analyzing the system on a sphere or torus (or any topologically closed surface) may well be insufficient for comparison to real experimental conditions. In any case, there is plenty of room for additional numerical analysis of $\nu = 12/5$, and the work done so far certainly does not exclude our states as a viable alternative to the Abelian or $k=3$ RR states at $\nu=12/5$.

\section{Comparison to Experimental Studies}

While an intuitively appealing physical picture and extensive numerical analyses are important and helpful, one should keep in mind that the main focus of arguments in support or opposition of any proposal will undoubtedly turn back to empirical data, once more is obtained from the experiments currently being pursued. Several types of experiments are needed to determine the true nature of the observed FQH states. Interferometry experiments that probe braiding statistics can provide a ``smoking gun,'' to unambiguously identify states as non-Abelian~\cite{Chamon97,Fradkin98,DasSarma05,Stern06a,Bonderson06a}, and can be used to distinguish between certain classes of topological order~\cite{Bonderson06b}, for example between Abelian, Ising, and Fibonacci (i.e. between HH, MR/BS, and $k=3$ RR) type states. However, certain ``similar'' states, such as those with Ising fusion rules in the non-Abelian sector, cannot be distinguished from braiding alone. Experiments that probe scaling behavior~\cite{Wen92b,Fendley06a,Fendley07a} or thermal Hall conductance~\cite{Kane97} should, in principle, be able to distinguish between such ``similar'' states. Though the necessary evidence for a complete characterization of all the second Landau level states is currently lacking, rapid progress is being made on the experimental front.

The decisive experimental evidence that does exist at present is in complete agreement with our proposed states. This essentially means our states occur at the observed filling fractions, and have fundamental quasihole charges that match the measured values in Ref.~\onlinecite{Dolev08}. This is notable for the plateau at filling fraction $\nu=7/3$ and $8/3$, since our states are presently the only proposed spin-polarized, non-Abelian candidates for these plateaus with $e/3$ fundamental quasihole charge. Distinguishing states by their fundamental quasihole charge is more difficult for $\nu=12/5$, as the Abelian hierarchy state, the $k=3$ RR state, and our proposed states all have the same $e/5$ fundamental quasihole charge.

\begin{table}
\[
\begin{array}{|c|r|r|r|r|r|}
\hline
\nu&\phantom{\frac{1}{1_{j}}}\frac{7}{3}&\frac{12}{5}&\frac{5}{2}&\frac{8}{3}&\frac{14}{5}\\
\hline
\Delta_{\cite{Pan99}}           &          100 &          \star &     110 &      55 &         \\
\Delta_{\cite{Eisenstein02}}    &       \star  &  \phantom{888} &     310 &   \star &  \star  \\
\Delta_{\cite{Xia04}}           &      \sim 600&              70&   \star &   \star &  \star  \\
\Delta_{\cite{Choi07}}          &           584&         \star  &      544&      562&    252  \\
\Delta^{\prime}_{\cite{Choi07}} &           206&                &      272&      150& \leq 60 \\
\Delta_{\cite{Miller07a}}       &           110&                &      130&       60&         \\
\Delta_{\cite{Pan08}}           &           590&         \star  &      450&      290&  \star  \\
\Delta_{\cite{Dean08}}          &           225&                &      262&       64&    149  \\
\hline
\end{array}
\]
\caption{Experimentally observed filling fractions and excitation gaps (in mK) of fractional quantum Hall states in the second Landau level. The subscript on $\Delta$  is the number of the reference where the corresponding gap values were reported (Ref.~\onlinecite{Choi07} reported data for two different samples). A $\star$ indicates an observed plateau, but no gap value reported.}
\label{Table:gaps}
\end{table}

On a more qualitative level, one can argue that the order of appearance and relative strengths of measured energy gaps in the second Landau level (see Table~\ref{Table:gaps}) put constraints on what theories should be proposed, and provide support for our hierarchy picture in the same way that they do for the Abelian hierarchy in the lowest Landau level. In this vein, one may argue that the observation~\cite{Xia04,Pan08} of a strong feature (or more optimistically, a FQH plateau) at $\nu=21/8$ supports our proposal, as such a state emerges naturally from our $\nu = 12/5$ state by adding one additional layer of hierarchy (or alternatively by filling one additional Landau level of composite fermions). It is notable that the $\nu=5/2$ state, which we use as the parent state for our hierarchy, usually has the largest gap, while states that occur at higher hierarchical levels have decreasing gaps. An interesting empirical property of the second Landau level is the apparent lack of particle-hole symmetry. Even in the case $\nu=7/3$ and $8/3$, where the filling fractions are particle-hole conjugate, the gaps (in all but one sample) have a large disparity $\Delta_{7/3} \sim 2 \Delta_{8/3}$, so it seems likely that these states are not particle-hole conjugates of each other (though possibly depending on the sample and experimental conditions). In particular, this suggests that (at least) one of them is not a standard Abelian state, and a natural guess would be that it is thus a non-Abelian state.

\section{Conclusion}

We have shown how to perform general hierarchical constructions of FQH states. Applying this framework to the $\nu=5/2$ Moore--Read state, we have constructed hierarchies of states (and their corresponding trial wavefunctions) that we propose as candidates for all the observed FQH states (as well as features suggestive of developing FQH states) in the second Landau level. The quasiparticle excitations of these proposed states all have non-Abelian statistics of the Ising type. The resulting hierarchy provides a picture in which $p$-wave pairing characteristic of the MR state is the ubiquitous mechanism that generates non-Abelian statistics in FQH states. These proposed states are consistent with all the presently available experimental evidence.

This hierarchical framework can also be applied to any other candidate for the $\nu = 5/2$ state and similarly produce new candidate states for all the observed second Landau level plateaus. In particular, one can build a hierarchy on the $ \text{SU} \left(2\right)_{2} \times \text{U}\left(1\right)_{2}$ NAF state~\cite{Blok92}, or similarly constructed states given by $ \text{SO} \left(5\right)_{1} \times \text{U}\left(1\right)_{2}$ or $ \text{SO} \left(7\right)_{1} \times \text{U}\left(1\right)_{2}$. We note that Ising, $ \text{SU} \left(2\right)_{2}$, $ \text{SO} \left(5\right)_{1}$, and  $ \text{SO} \left(7\right)_{1}$ all have the same fusion algebra, and all have a fermionic field (the equivalent of the $\psi$) assigned to the electron in such states. One can also use versions of all these states in which the non-Abelian sectors occur in the opposite chirality, i.e. $\overline{ \text{Ising} } \times \text{U}\left(1\right)_{2}$, $ \overline{ \text{SU} \left(2\right)_{2} } \times \text{U}\left(1\right)_{2}$, etc. Altogether, this provides (at least) sixteen distinct sets of candidate states (including the particle-hole conjugates) covering all observed filling fractions of the second Landau level. Though these other candidates are mostly unlikely, the point is that one may apply a hierarchical construction to whatever state is the correct description of $\nu=5/2$, and produce candidates for all the other observed second Landau level states. Also, for a given filling fraction, the states built from theories with the Ising fusion algebra in their non-Abelian sector will all have the same fundamental quasihole charges, but slightly different braiding statistics and scaling behavior.

It is straightforward to produce similar hierarchies over other non-Abelian states, such as the RR states; however it is also less fruitful in that the observed second Landau level filling fractions are not naturally produced, nor does it provide a conceptual picture of second Landau level FQH physics that is as appealing.

Finally, we remark that while non-Abelian FQH states are of great interest in their own right, their study has an additional impetus provided by their prospective application in topological quantum computation~\cite{Kitaev03,Freedman02a}. We recall that the quasiparticle excitations of the states we have proposed here will have the same braiding statistics as the parent state at $\nu=5/2$, up to Abelian factors coming from the hierarchy. In particular, this means that neither the hierarchy built on the MR state, nor those built upon any of the other current candidate states for $\nu=5/2$ will have computationally universal non-Abelian braiding statistics [see Eqs.~(\ref{eq:Isingfusion},\ref{eq:Isingbraiding})]. In order to achieve universal quantum computation with such states, quasiparticle braiding~\footnote{We note that while braiding transformations are used to generate computational gates in topological quantum computation, strictly speaking one does not actually need to physically move the computational anyons, as it was shown in Refs.~\onlinecite{Bonderson08a,Bonderson08b} that a series of topological charge measurements could be use to mimic the braiding transformations of anyons.} must be supplemented by operations that are topologically unprotected~\cite{Bravyi06} or involve changing the topology of the system~\cite{Freedman06a,FNW05b}. On the other hand, the non-Abelian statistics of the $k=3$ RR state is described by Fibonacci anyons, which are known to have computationally universal braiding~\cite{Freedman02b}. Thus, it is of great interest for topological quantum computation whether the $\nu=12/5$ state is RR, BS, or HH (or something else entirely) as the outcome to this question will have fundamental consequences for its implementation in FQH systems.

\begin{acknowledgments}
We thank Jim Eisenstein, Adrian Feiguin, Woowon Kang, Chetan Nayak, Wei Pan, Ed Rezayi, Kirill Shtengel, and Steve Simon for
illuminating discussions. We also acknowledge the hospitality of the IQI, Microsoft Station Q, and the Aspen Center for Physics. This work
was supported in part by the NSF under Grant No.~PHY-0456720 and the ARO
under Contract No.~W911NF-05-1-0294.
\end{acknowledgments}

\end{document}